\documentclass{epl}

\usepackage{amsmath}
\newcommand{\Eref}[1]{eq.~(\ref{#1})}
\newcommand{\Fref}[1]{fig.~\ref{#1}}
\title{Comment on ``Anisotropic $s$-wave superconductivity: comparison
with experiments on MgB$_2$'' [A.~I.~Posazhennikova, T.~Dahm and
K.~Maki, cond-mat/0204272; submitted to \textit{Europhys. Lett.}]}
\shorttitle{Comment on ``Anisotropic $s$-wave superconductivity\dots''}

\author{T. M. Mishonov\inst{1,2} \and E. S. Penev\inst{2} \and J. O. Indekeu\inst{1}}

\institute{
  \inst{1} Laboratorium voor Vaste-Stoffysica en Magnetisme,
    Katholieke Universiteit Leuven -
    Celestijnenlaan 200 D, B-3001 Leuven, Belgium\\
  \inst{2} Department of Theoretical Physics, Faculty of Physics,
    Sofia University ``St. Kliment Ohridski'' -\\
    5 J. Bourchier Blvd., Bg-1164 Sofia, Bulgaria
}


\begin{document}

\maketitle

\begin{abstract}
An analytical result for the renormalization of the jump of the heat
capacity $\Delta C/C_N$ by the anisotropy of the order parameter is
derived within the framework of the very recent model proposed by
Posazhennikova, Dahm and Maki [cond-mat/0204272 submitted to
\textit{Europhys. Lett.}], for both oblate and prolate anisotropy. The
graph of $\Delta C/C_N$ versus the ratio of the gaps on the equator
and the pole of the Fermi surface, $\Delta_e/\Delta_p$, allows a
direct determination of the gap anisotropy parameter
$\Delta_e/\Delta_p$ by fitting data from specific heat measurements
$\Delta C/C_N$. Using the experimental value $\Delta
C/C_N=0.82\pm10\%$ by Wang, Plackowski, and Junod [\textit{Physica C} 
\textbf{355} (2001) 179] we find $\Delta_e/\Delta_p\approx4.0$.
\end{abstract}

In a very recent e-print Posazhennikova, Dahm and Maki~\cite{Dahm:02}
discuss a model for the gap anisotropy in MgB$_2$, a material which
has attracted a lot of attention from condensed matter physicists in
the past two years.  A central issue in this work~\cite{Dahm:02} is to
propose an analytic model for analyzing thermodynamic
behavior. Assuming a spherical Fermi surface, a simple gap anisotropy
function is suggested, $\Delta ({\bf k})=\Delta_e/\sqrt{1+ A z^2}$,
where $z= \cos\theta$, and $\theta$ is the polar angle.  This model
leads to useful results for the temperature dependence of the upper
critical field $H_{c2}$ and of the specific heat, which can be fitted
to the experimental data, thereby determining the optimal anisotropy
parameter $A$. Note that $A=(\Delta_e/\Delta_p)^2-1$, with $\Delta_p =
\Delta (z=1)$ and $\Delta_e = \Delta (z=0)$, and the gap ratio is
parameterized as $\Delta_e/\Delta_p=\sqrt{1+A}>0$.

The aim of the present Comment is to provide a convenient analytical
expression giving the possibility for determining $\Delta_e/\Delta_p$
from the available data for the jump of the specific heat~\cite{Wang:01}.
For the latter we derive the explicit formula
\begin{equation}
 \frac{\Delta C}{C_N} = \frac{12}{7\zeta(3)}\,\frac{1}{\beta_{\Delta}},
       \mbox{ where }
       \frac{1}{\beta_{\Delta}}=
       \frac{\langle\left|\Delta_\mathbf{k}\right|^2\rangle^2}
       {\langle1\rangle\langle\left|\Delta_\mathbf{k}\right|^4\rangle},
 \quad
 \langle f(\mathbf{k})\rangle=2\sum_b\int
 \delta(\varepsilon_{b,\mathbf{k}}-E_\mathrm{F})f(\mathbf{k})\,
 \frac{d\mathbf{k}}{(2\pi)^3},
\label{general}
\end{equation}
$E_{\mathrm{F}}$ is the Fermi energy, $\varepsilon_{b,\mathbf{k}}$ are
the band energies, $\langle1\rangle$ is the density of states, $\zeta$
is the Riemann zeta function, $\beta_{\Delta}$ is analogous to Abrikosov's
parameter $\beta_A$~\cite{Abrikosov:88}, and $12/7\zeta(3)=
1.42613\dots$ is the sacramental BCS ratio.
\begin{figure}[t]
\onefigure[width=0.5\textwidth]{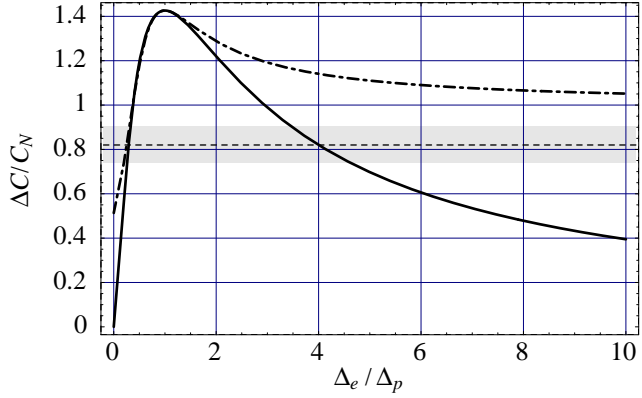}
\caption{Jump of the specific heat $\Delta C/C_N$ versus the
``equatorial-to-polar'' gap ratio $\Delta_e/\Delta_p$. For a given
$\Delta C/C_N$ value we have oblate $\Delta_e/\Delta_p>1$ and prolate
$\Delta_e/\Delta_p<1$ solutions. The solid line gives our present
analytical result \Eref{eq:oblate} for the model by Posazhennikova,
Dahm and Maki~\cite{Dahm:02}. The dash-dotted line is our analytical
solution~\cite{MPI:02} for the model by Haas and
Maki~\cite{Haas:01}. The dashed line is the jump
ratio $\Delta C/C_N=0.82\pm10\%$ measured by Wang, Plackowski, and
Junod~\cite{Wang:01}, with the shaded area showing the experimental error
bar.
\label{fig:1}
}
\end{figure}
Then following the weak-coupling BCS approach~\cite{Dahm:02,Haas:01}
we derived the explicit analytic expressions
valid for $A>0,$ and $-1<A<0$, respectively
\newcommand{\D}{\displaystyle}
\begin{equation}
\frac{\Delta C(A)}{C_N} =
\begin{cases}
  \frac{\D 12}{\D 7\zeta(3)}
      \frac{\D 2(1+b^2)\left(\arctan b\right)^2}
           {\D b^2+b(1+b^2)\arctan b},&
 b = \sqrt{A}=\sqrt{\left(\frac{\Delta_e}{\Delta_p}\right)^2-1},
\label{eq:oblate}\\[0.4cm]
 \frac{\D 12}{\D 7\zeta(3)}
      \frac{\D 2(1-p^2)\left(\tanh^{-1} p\right)^2}
           {\D p^2+p(1-p^2)\tanh^{-1} p}, &
 p = i b = \sqrt{-A}=\sqrt{1-\left(\frac{\Delta_e}{\Delta_p}\right)^2}.
\end{cases}
\end{equation}
For a given specific heat jump, this expression leads to \emph{two}
solutions (oblate, $\Delta_e/\Delta_p>1,$ and prolate,
$\Delta_e/\Delta_p<1$). The relevant example is shown in
\Fref{fig:1}; the function $\Delta C(A)/C_N$
is tabulated in ref.~\cite{Dahm:02}.
The analysis of the angular dependence of
$H_{c2}$~\cite{Xu:01,Angst:02} performed in the commented
paper~\cite{Dahm:02} unambiguously demonstrates that one has to
analyze only the ``oblate'' case. Thereby the experimentally reported
value in ref.~\cite{Wang:01} $\Delta C/C_N=0.82\pm10\%$ gives
$A\approx 16$ and $\Delta_e/\Delta_p\approx b \approx 4.0\pm 10\%$.
For this significant anisotropy, the ``distribution'' of Cooper pairs
%
$
\langle\left|\Delta_\mathbf{k}\right|^2\rangle
 \propto 1/[k_z^2+(k_{\mathrm{F}}/b)^2]
$
%
has a maximum at $k_z=0$. This general qualitative
conclusion is in agreement with the hints from band calculations
that the maximal order parameter is concentrated in an almost two-dimensional
electron band, but all bands $\varepsilon_{b,\mathbf{k}}$
take part in the normal specific heat per unit cell
$C_N=(\pi^2/3)k_B^2T\langle1\rangle$.

For the two-band model, advocated for the first time for MgB$_2$ in
ref.~\cite{Shulga:01}, \Eref{general} gives (to within a typographical
correction) the result by Moskalenko~\cite{Moskalenko:59}
\begin{equation}
\label{eq:2band}
 \frac{\Delta C}{C_N} = \frac{12}{7\zeta(3)}\,
 \frac{(|\Delta_1|^2 \rho_1+ |\Delta_2|^2 \rho_2)^2}
 {(\rho_1+\rho_2)(|\Delta_1|^4 \rho_1+ |\Delta_2|^4 \rho_2)}
 =1.426\,\frac{[z^2 x + (1-x)]^2}{z^4x+(1-x)},\quad
 \mbox{ where }
 z=\frac{\Delta_1}{\Delta_2},\;
 x=\frac{\rho_1}{\rho_1+\rho_2},
\end{equation}
and $\rho_1$ and $\rho_2$ are the densities of states for the two
bands. Taking for an illustration $x=0.515$ and $\Delta C/C_N=0.82,$
\Eref{eq:2band} gives $\Delta_1/\Delta_2\approx 4.0$ in agreement with
$\Delta_\mathrm{e}/\Delta_\mathrm{p}\approx 4.0$ obtained using
\Eref{eq:oblate}. Thus the gap ratios are model-independent. For a
survey on a set of parameters see Table~I in
ref.~\cite{Bouquet:01}. Certainly the jump of the heat capacity alone
cannot be an arbiter for the validity of any model, so subtleties,
e.g., related to strong coupling effects and other anisotropies, can
be hidden in the parameters spread in the table mentioned.

\begin{acknowledgements}
We are thankful to S.-L.~Drechsler for bringing
ref.~\cite{Moskalenko:59} to our attention.  This work was supported
by the Flemish programmes IUAP and GOA.
\end{acknowledgements}


\begin{thebibliography}{9}

\bibitem{Dahm:02}
  \Name{Posazhennikova A.~I.,  Dahm T. \and Maki K.}
  cond-mat/0204272, submitted to \textit{Europhys. Lett.}.

\bibitem{Wang:01}
 \Name{Wang Y., Plackowski T. \and Junod A.}
    \REVIEW{Physica C}{355}{2001}{179}.


\bibitem{Abrikosov:88}
  \Name{Abrikosov~A.~A.}
  \Book{Fundamentals of the Theory of Metals}
  \Publ{North-Holland, Amsterdam} \Year{1988}, eq.~(18.25).

\bibitem{Haas:01}
 \Name{Haas~S.\and Maki~K.}
    \REVIEW{Phys. Rev. B}{65}{2001}{020502(R)}.

\bibitem{MPI:02}
 \Name{Mishonov~T.~M., Penev~E.~S. \and Indekeu~J.~O.}
    cond-mat/0204545, submitted to \textit{Phys. Rev.~B.}

\bibitem{Xu:01}
 \Name{Xu~M., Kitazawa~H., Takano~Y., Ye~J., Nishina~K.,
    Abe~H., Matsushita~A., Tsujii~N. \and Kido~G.}
    \REVIEW{Appl. Phys. Lett.}{79}{2001}{2779}.

\bibitem{Angst:02}
 \Name{Angst~M., Puzniak~R., Wisniewski~A., Jun~J.,
    Kazakov S.~M., Karpinski J., Roos J. \and Keller~H.}
    \REVIEW{Phys. Rev. Lett.}{88}{2002}{167004}.
\bibitem{Shulga:01} \Name{Shulga~S.~V., Drechsler~S.-L., Eschrig~H.,
Rosner~H. \and Pickett~W.} \textit{``The upper critical field problem
in MgB$_2$''}, cond-mat0103154.
\bibitem{Moskalenko:59} \Name{Moskalenko~V.~A.}  \REVIEW{Fiz. Met. i
Metalloved.}{8}{1959}{503} [\textit{Phys. Met. Metallogr. (USSR)}];
\Name{Moskalenko~V.~A. \and Palistrant~M.~E.} ``Theory of pure
two-band superconductors'' in \textit{Statistical Physics and Quantum
Field Theory -- in memoriam to S.~V.~Tyablikov} (in Russian:
\textit{Staitisticheskaya fizika i kvantovaya teoriya polya}),
ed. N.~N.~Bogoliubov (Moskow, Nauka, 1973), p.~226.

\bibitem{Bouquet:01}
 \Name{Bouquet~F., Wang~Y., Fisher~R.~A., Hinks~D.~G.,
Jorgensen~J.~D., Junod~A. \and Phillips~N.~E.}  
 \REVIEW{Europhys. Lett.}{56}{2001}{856}.

\end{thebibliography}
\end{document}